# Accessibilité aux informations textuelles et visuelles sur les sites web pour les personnes avec une déficience visuelle

Katerine ROMEO, Edwige PISSALOUX
Laboratoire LITIS/CNRS FR 3638
Université de Rouen Normandie
Saint Etienne du Rouvray, France
katerine.romeo@univ-rouen.fr

Frédéric SERIN
Laboratoire LITIS/CNRS FR 3638
Université du Havre Normandie
Le Havre, France

*Abstract*—L'accès aux informations numériques textuelles et visuelles pour les personnes ayant une déficience visuelle (PPIV) devient très difficile avec les lecteurs d'écran qui ne sont pas adaptés aux différents sites sur le web. Ce papier analyse l'utilisation des différentes technologies pour accéder au contenu numérique et établir des améliorations de recommandations existantes pour la conception des sites web accessibles à tous. Les résultats d'évaluation préliminaire avec les PPIV du site web de notre projet ACCESSPACE réalisé en tenant compte des améliorations proposées confirment leur pertinence.

## I. Introduction

Nous sommes tous « connectés » et sollicités en permanence pour rechercher des informations sur le web, pour nous inscrire sur différents sites, pour remplir des formulaires en ligne, pour envoyer et recevoir des courriels, pour utiliser différentes applications mobiles, etc. Ces actions font partie intégrante de la société numérique. Or cette société est multimodale avec la prédominance de la vision et de la perception visuelle. Cette prédominance constitue un sérieux frein à l'accès à l'information numérique et à l'intégration de tous à cette société numérique, notamment des personnes présentant une incapacité visuelle (PPIV) et des séniors ; elle crée « l'illettrisme numérique » de ces membres de notre société.

Cette situation d'exclusion devient préoccupante car, selon l'OMS (Organisation Mondiale de la Santé) 285 millions de personnes souffrent de déficience visuelle dans le monde [1]. Parmi les PPIV répertoriées, 4,5 % environ ont une déficience de perception de couleur (la difficulté de distinction de teinte des couleurs et de sensibilité à la luminosité), 4 % de personnes ont une vision tubulaire ou un champ de vue réduit ou encore des problèmes d'acuité visuelle, et 0,6 % sont des non-voyants [2]. Ces maladies impactent directement la localisation de l'information sur les sites internet ainsi que l'acquisition et la compréhension des connaissances qui y sont attachées.

Plusieurs causes peuvent être à l'origine des difficultés d'accès à l'information contenue sur les sites web ; elles peuvent être liées aux sites et/ou aux technologies d'accès aux sites, présentes sur le marché, tant logicielles (ex. lecteurs d'écran) que matérielles (ex. les interfaces à stimulation tactile). Aussi, la conception et la réalisation des sites internet accessibles à tous, accessibles de façon intuitive, exigent une approche holistique et une expérimentation avec « un laboratoire vivant » (living lab).

Ce papier propose une approche holistique du problème d'accessibilité aux sites internet. Après une brève analyse des mécanismes cognitifs de lecture que l'on met en place lors de son apprentissage (II), on analyse l'état de l'art des technologies actuelles (III). Ces deux sections permettent d'établir des améliorations de recommandations existantes pour la conception des sites web accessibles à tous (section IV). Finalement la section (V) présente : (1) une méthode de conception et d'évaluation d'accessibilité de sites web, et (2) les résultats préliminaires d'évaluation d'un site web réalisé pour le projet ACCESSPACE. Finalement, la section (VI) résume les principaux résultats de cette recherche, et propose leur extension.

## II. Mecanismes cognitifs d'exploration d'un document numérique

L'apprentissage de lecture de personnes saines et de déficients visuels est basé sur les mêmes mécanismes qui ont été traduits plus ou moins efficacement par les technologies d'assistance de lecture. Cette section rappelle les éléments usuels constituant une page et un document à lire (A), puis les principaux mécanismes naturels d'exploration d'un document lu (B), des supports de représentation de l'information (C), et leurs inconvénients (D).

### A. Elements usuels constituant un document numérique

La structuration d'un document numérique essaie de mettre en œuvre les lois de perception (théorie de Gestalt [12]). Cette théorie stipule que le lecteur utilise ses capacités naturelles d'organisation de l'information par regroupement et en distinguant les différents éléments. La perception visuo-spatiale utilise le principe de similarité, la hiérarchie visuelle et la complexité hiérarchique. L'espace est organisé en utilisant les attributs pour la ségrégation visuelle comme des éléments graphiques ou des effets de contraste.

Ces principes ont été implantés dans un document numérique en l'organisant en deux parties obligatoires. La première partie (l'en-tête) est constituée des métadonnées, c'est-à-dire les informations qui identifient le document et qui caractérisent son contenu. Cette partie permet donc l'indexation de tout document électronique. La deuxième partie est le plan structurel qui présente l'organisation logique du document qui sous-tend le discours, les propos développés. Il comprend des titres, des chapitres, des sections, des sous-sections, des blocs de texte, des blocs d'images, des tableaux, une bibliographie, etc. [3] [4].

Un document HTML est un exemple-type d'un document électronique. Il est constitué d'un en-tête qui précise le titre de la page, les métadonnées de description et de mots clés. Certaines de ces informations sont utilisées pour le référencement de la page web. Le corps de la page (appelé body) est constitué d'éléments sémantiques comme en-tête, menu de navigation, sections, articles, pied de page, etc. Chaque élément d'un document HTML est encadré par une balise qui indique la fonction de cet élément. Un élément de la page courante peut contenir d'autres éléments comme paragraphes, images, vidéos, formulaires etc. Ainsi, les balises sont imbriquées.

Trois caractéristiques syntaxiques inhérentes facilitent l'accès au contenu [6] et attirent l'attention sur : la visibilité (ou le contraste), la lisibilité et l'attention visuelle, qui permet de repérer le contenu d'un document et d'appréhender son sens.

*B. Lecture naturelle d'un document numérique*

Nous utilisons différentes stratégies cognitives pour lire un document. Une stratégie est une démarche consciente mise en place pour comprendre le document ; elle intègre une suite de processus cognitifs élémentaires (Fig. 1) [5].

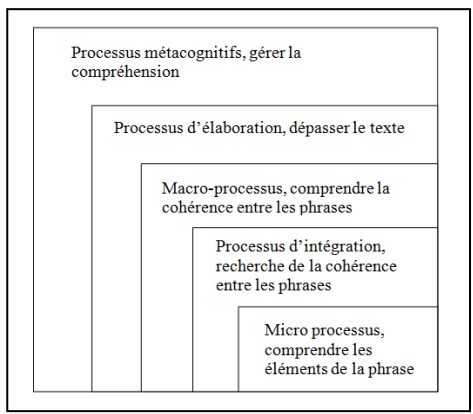

Fig. 1 Processus cognitifs de base connus sous-tendant la lecture [5]

Le processus de lecture et sa vitesse varie avec le but de la lecture. Comme le montre le tableau 1, la lecture pour apprendre et pour mémoriser prend trois à quatre fois plus de temps qu'un balayage seul d'une page.

| Activité | Fonction | Vitesse de lecture |
|---|---|---|
| Balayage | Rechercher un mot rapidement | 600 mots/mn |
| Ecrémage | Rechercher un contenu rapidement | 450 mots/mn |
| Lecture normale | Lecture silencieuse | 300 mots/mn |
| Lire pour apprendre | Lire pour acquérir un nouveau contenu | 200 mots/mn |
| Lire pour mémoriser | Lire pour mémoriser un texte | 138 mots/mn |

Tableau 1. Différents vitesses de lecture en fonction de l'activité [6] [7]

La vitesse de lecture est une variable personnelle ; on a estimé qu'un lecteur rapide lit (via les saccades) plus de 500 mots par minute, qu'un lecteur moyen lit entre 200 et 250 mots par minute et qu'un lecteur lent lit moins de 100 mots par minute [6]. Cette vitesse dépend aussi du type de document à lire et de son support matériel: linéaire (une page web, un livre) (1) ou non-linéaire (plusieurs pages web) (2) car les processus cognitifs mis en jeu ne sont pas les mêmes.

*1) Lecture naturelle d'un document linéaire:* La lecture d'un document linéaire, par exemple, un livre, se fait de façon séquentielle ; les mots sont identifiés et la position de certains mots importants dans la page est mémorisée. Ce processus permet des retours en arrière à une place fixe pour une meilleure assimilation de certaines informations. Le contexte reste donc stable, et les processus d'appréhension et de mémorisation de l'information sont renforcés.

Lorsqu'on lit un texte sur une page de web, ce processus de mémorisation spatiale de mots est dégradé avec le déroulement (scrolling) vertical, voire avec les mises à jour instantanées de page (par exemple les documents partagés, travail collaboratif): le lecteur est face à une instabilité spatiale et perd ses repères spatiaux.

*2) Navigation naturelle dans un document non-linéaire:* Les documents non-linéaires comprennent des hypertextes qui détournent l'attention du lecteur à cause du contexte élargi de la recherche. La consultation de plusieurs pages en parallèle, tout en enrichissant les connaissances sur le sujet, peuvent néanmoins amener la confusion avec des répétitions. Dans le but de compréhension d'un document non-linéaire, le lecteur doit se construire une représentation mentale du contenu. Cette représentation est complétée ensuite par les recherches effectuées dans le document. Ainsi l'image mentale du contenu évolue en une structure arborescente car c'est une structure algorithmique habituellement mise en œuvre par les réalisateurs de sites, et exige des nouveaux processus de navigation dans une arborescence. Aussi, cette activité de navigation dans un document non-linéaire utilise différents processus cognitifs : de navigation, de sélection de l'information, de traitement spatial du document et de la construction de la représentation de l'information dans le document [8]. A cela s'ajoute le partage stratégique de l'attention nécessaire à l'analyse de plusieurs médias, textes, images, schémas, vidéos ou animations, ce qui entraîne une charge cognitive importante.

La principale difficulté de lecture d'un document non-linéaire électronique, réside dans le fait qu'à chaque collecte d'information sur internet une nouvelle facette de l'information peut apparaître (une instabilité de connaissance). Pour stabiliser la recherche d'information et écarter les informations non pertinentes un raisonnement spécifique devrait être mis en œuvre, comme le test d'hypothèses [9]. Si le but à atteindre est défini, chaque hypothèse (découverte d'une nouvelle facette de l'information) est successivement examinée, acceptée ou rejetée [10].

*3) Accès au document et la déficience visuelle :* Force est de constater que tous les éléments d'accès et d'exploration d'un document restent dans le domaine du visuel et sont non accessibles aux personnes présentant une déficience visuelle (PPIV) qui sont privées de la perception visuo-spatiale. La navigation dans un document ou sur internet pour la PPIV se passe différemment car avec le lecteur d'écran elle parcourt de façon séquentielle tout le document. Elle s'attend à trouver une structure de document linéaire ; e.g. pour un site web, le menu pour la navigation doit être toujours situé tout en haut à gauche de la page d'accueil. C'est le point de départ pour l'exploration du site qui devrait être accessible en premier avant toute autre information. Le menu de navigation devrait se trouver aussi dans le plan du site car les PPIV ont l'habitude de le consulter pour trouver les pages qui les intéressent. Ces hypothèses sont reprises par les lecteurs d'écran et renforcent le processus cognitif de recherche d'information des PPIV basé sur un balayage séquentiel d'un document (comme le balayage vidéo).

Pour les parties saillantes d'un document (comme des animations, des vidéos et des images), les PPIV peuvent consulter le texte explicatif pour les images, le sous-titrage et l'audiodescription pour les animations et les vidéos, à condition que ces moyens alternatifs aient été ajoutés. Mais ces opérations additionnelles exigent une concentration, une imagination et le temps.

Le traitement spatial du document pour les PPIV se réduit donc à une structure séquentielle et la représentation de l'information qui dépend étroitement du but recherché. Une PPIV ayant de l'expérience sur internet survole les titres qui ne correspondent pas à ce qu'il recherche, mais peut être très vite perdue si de nouvelles fenêtres s'ouvrent sans prévenir et l'empêchent de revenir au point de départ.

*C. Supports de représentation de l'information*

La représentation structurelle d'un document HTML est organisée par le Modèle Objet du Document (DOM) qui est une API (interface de programmation). Le DOM permet d'accéder aux éléments de la page web pour modifier le document et sa présentation visuelle en reliant les pages web aux scripts et langages de programmation. Les balises HTML décrivent la structure d'un document web et indiquent au navigateur comment afficher le document et les différents médias comme des images, des vidéos, fichiers audio, etc. La représentation structurelle de la page web est organisée de façon arborescente. Le navigateur parcourt cet arbre au chargement de la page web et affiche les différents éléments selon les indications des balises et du style de la page web.

Les balises comme en-tête (header), pied-de-page (footer), bloc vertical de côté (aside) devraient être situés à leur place de définition. La figure 2 montre l'organisation spatiale attendue par rapport à la définition des balises. Le menu (nav), devrait se placer le plus haut possible. Le plan du site devrait se situer dans le pied de page (footer). La figure 3 montre l'arborescence de la structure de la page web de la figure 2.

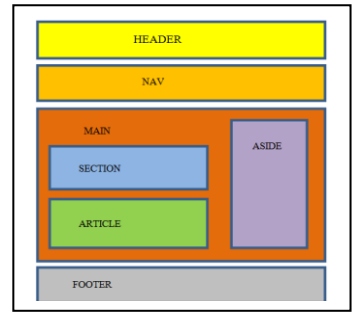

Fig. 2 Structure type d'une page web et son organisation spatiale

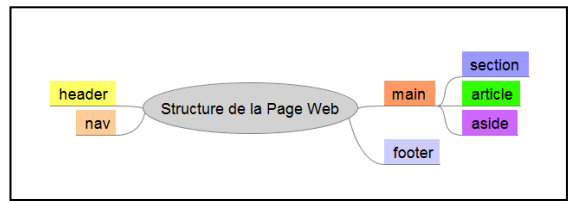

Fig. 3 Arborescence de la structure de la page web de la Fig. 2

*D. Inconvénients de certains supports de représentation de l'information*

Les personnes ayant une déficience visuelle, tout comme les personnes avec une vue correcte, cherchent une structure logique dans un document pour se construire une représentation correcte de l'information. L'organisation non explicite d'une page web où l'information recherchée est difficile à être localisée (car les balises HTML utilisées ne correspondent pas à leur fonction) ou l'organisation spatiale non habituelle des éléments du document peuvent désorienter les lecteurs. Les PPIV sont démunies devant des informations accessibles seulement visuellement comme la présence des informations 2D (ex. des tableaux, des images, des captchas graphiques, etc.). On peut citer aussi les formulaires exigeant des informations obligatoires à rentrer dans les champs indiqués en couleur, ou les formules mathématiques.

Les inconvénients des lecteurs d'écrans sont surtout liés à une représentation interne non standard d'un document ce qui induit un algorithme de navigation dans la page et dans le document qui varie d'un lecteur d'écran à un autre. Les PPIV sont amenées à effectuer des opérations élémentaires inutiles ne permettant pas d'appréhender rapidement et de façon intuitive l'organisation logique du document (par exemple le retour à la page précédente entraîne souvent sa relecture dès la première ligne de la page initiale au lieu de la place où la page suivante a été appelée, etc.), et entraîne une surcharge cognitive [15], [16], [17], [18].

Des solutions sont proposées pour les personnes ayant des problèmes de distinction de couleur. Dès l'entrée sur le site, un réglage de couleur donne la possibilité d'augmenter le contraste ou sur demande des malvoyants, des traitements sur les images en les rendant monochrome ou en niveaux de gris (Fig. 4). Ces images peuvent être affichées sur un support tactile (cf. III.B).

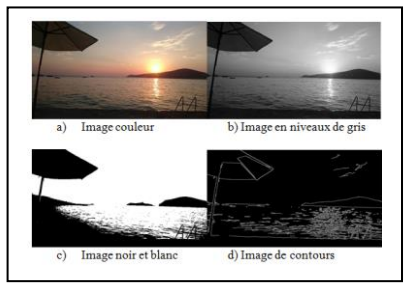

Fig. 4 Images en niveaux de gris (b), en noir et blanc (c) ou avec des contours (d) obtenues à partir d'une image couleur (a)

### III. TECHNOLOGIES ACTUELLES D'ACCES A L'INFORMATION POUR LES PERSONNES AVEC UNE DEFICIENCE VISUELLE

Pour réduire le problème de l'accessibilité des documents web il faut disposer d'un ensemble de moyens (audio et/ou à stimulation tactile) qui peuvent être utilisés comme interfaces pour tous, y compris les déficients visuels. Il s'agit d'outils de substitution : visuo-auditive et/ou visuo-tactile.

#### A. Outils de substitution visuo-auditive

Les lecteurs d'écran sont aujourd'hui des outils les plus représentatifs (et populaires) de substitution visuo-auditive.

Ils fournissent souvent deux fonctions de base :

*1) Annonce le contenu de la page courante*

Les premiers langages du web ont été conçus en réponse au besoin d'indiquer au navigateur la structure du document et comment l'afficher. Le langage de balisage HTML (Hypertext Markup language, 1991) avait les premières balises comme paragraphe, image, lien etc.

Le lecteur d'écran JAWS (Job Access With Speech, 1989) accède au DOM (Document Object Model) et utilise les interfaces de programmation du navigateur pour avoir les informations sur les objets, leur rôle, leur nom et leur état [11] afin de les annoncer au moment du parcours de la page. Un autre lecteur d'écran pour Windows est NVDA (Non Visual Desktop Access) développé en 2007 en open source.

Les lecteurs d'écran pour les déficients visuels utilisent les balises HTML; le contenu de la page est exploré séquentiellement en parcourant la liste de balises (en appuyant sur la touche « suivant », tab ou flèche). Les balises comme des titres, des liens ou des graphiques sont annoncées suivies du contenu de la balise.

*2) Restitution de l'information en synthèse vocale ou en Braille :*

Le lecteur d'écran NVDA comme JAWS, annonce le contenu d'une page par synthèse vocale ou en braille. Il est possible de régler la vitesse de lecture et choisir la voix (féminine ou masculine, avec un degré d'intonation métallique). Un temps d'adaptation est nécessaire car la voix étant monocorde et sans intonation, le sens des mots échappe parfois pendant l'écoute. Comme le nombre des PPIV-braillistes diminue, les lecteurs d'écran (comme VoiceOver ou TalkBack) deviennent de plus en plus accessibles et faciles à utiliser. De plus, les seniors n'ont pas toujours la possibilité d'apprendre le braille et se tournent vers une restitution des informations par synthèse vocale.

#### B. Outils de substitution visuo-tactile

Le contenu des sites web contient souvent des documents visuels avec des explications textuelles. Nous pensons que ces contenus numériques graphiques ou avec des images doivent être accessibles tactilement pour augmenter la compréhension et l'échange de l'information [13]. Il existe des *prototypes* de tablettes qui visent à produire avec des formes simples ces informations visuelles. Ces appareils de substitution visuo-tactiles utilisent différentes technologies. Par exemple, la Stimtac de l'Université Lille 1 ou la Hap2U présentent une surface avec une vibration et un coefficient de friction plus ou moins intense [19] ; le Tactonom [20] reproduit une image avec 10500 points tactiles métalliques qui se placent par attraction magnétique sur une grille de taille A4 ; le Graphiti [21] présente l'image avec des picots à plusieurs hauteurs.

### IV. RECOMMANDATIONS ET DEFIS DANS LA CONCEPTION DES SITES WEB

W3C (World Wide Web Consortium) a publié des recommandations pour rendre accessibles les sites web et les applications web à tous y compris les PPIV. En 1997, un groupe de travail nommé WAI (Web Accessibility Initiative) a été créé pour proposer des spécifications sur l'accessibilité du Web. Les normes proposées se nomment WAI-ARIA (WAI-Accessible Rich Interactive Applications) ; elles ajoutent des marqueurs de sémantique et de métadonnées aux balises HTML à destination des outils d'assistance technique comme les lecteurs d'écran. Actuellement HTML5 prend en compte la sémantique et certains attributs ARIA [22] ne sont plus utiles. Les contenus dynamiques, par contre, utilisant la souris et le contrôle des interfaces utilisateur de plus en plus complexes requièrent des définitions ARIA.

Par ailleurs, les règles d'accessibilité WCAG (Web Content Accessibility Guidelines) considèrent qu'un site web devrait avoir les quatre attributs suivants :

(1) être perceptible, c'est-à-dire que le contenu non textuel devrait avoir une alternative textuelle de description, le contraste des couleurs et la taille du texte devraient être réglables, les sons devraient avoir une différence de plus de 20 dB par rapport au fond sonore ;

(2) être utilisable avec la souris et le clavier pour tous les éléments de l'interface utilisateur et de navigation ;

(3) être compréhensible, pour le fonctionnement de l'interface utilisateur et pour l'information du contenu de la page web ;

(4) être robuste par rapport aux outils d'accessibilité et aux navigateurs.

En France, depuis 2009, le Référentiel Général d'Accessibilité pour les Administrations (RGAA) propose des règles d'accessibilité des services de communication publique en ligne. Ces règles sont basées sur le WCAG. L'association française BrailleNet édite le référentiel AccessiWeb qui analyse la question de présence et de pertinence de chaque critère d'accessibilité. On peut citer aussi Opquast qui a établi 226 critères pour un site de qualité et d'accessibilité.

## V. EVALUATION DE L'ACCESSIBILTE D'UN SITE : EXEMPLE.

Nous avons réalisé un site web pour l'un de nos projets de recherche, Accesspace (accesspace.univ-rouen.fr). Sa création a été l'occasion d'appréhender les difficultés rencontrées par des PPIV (Fig. 5). Ci-dessous sont présentés : le plan et le parcours du site avec un lecteur d'écran, le suivi de la mise en œuvre des règles d'accessibilité, le protocole de l'évaluation du site par les participants au test préliminaire ainsi que les résultats d'évaluation.

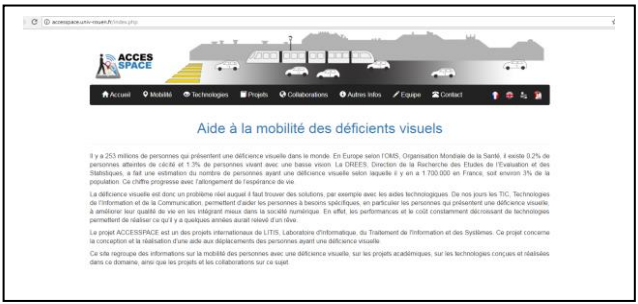

Fig. 5. Accesspace, le site évalué pour l'accessibilité

### A. Plan du site

Le plan du site permet d'avoir une vision globale du site web et des parcours de navigation. Il comprend les onglets de navigation entre les différentes pages, la navigation secondaire dans chaque page, c'est-à-dire le menu enfant de chaque élément sélectionné dans le menu principal, le pied de page avec un accès à des pages clés. La profondeur des pages est limitée à 3 pour faciliter la navigation de PPIVs (suite à leur recommandation).

### B. Parcours d'une page du site

Le parcours de chaque page du site avec le lecteur d'écran suit l'ordre de l'arborescence de la page en commençant à chaque fois par : l'intitulé de la barre de navigation, le logo, l'image de l'entête, le choix de langue, les titres des onglets du menu, le titre de la page, le texte, les images et les liens du contenu de la page, et du pied de page. La PPIV doit arrêter le lecteur à l'onglet ou le lien choisi en confirmant ce choix.

### C. Mise en oeuvre des règles d'accessibilité

En fonction du contenu des pages, les règles d'accessibilité matérialisées par des balises sont appliquées. Voici une liste non exhaustive de critères utilisés : les images et les vidéos sont indiquées par une alternative textuelle ; l'information n'est pas donnée uniquement par la couleur - des contrastes élevés sont utilisés pour la lisibilité du site ; les tableaux sont accessibles avec l'indication des en-têtes de colonnes et des lignes ainsi que le titre du contenu ; les liens, les abréviations, les listes et les boutons sont correctement balisés et indiqués ; la langue par défaut et les changements de langue sont identifiés ; la hiérarchie des titres est respectée ; pour les formulaires chaque champ est associé à son intitulé ; la description de chaque page est ajoutée.

### D. Protocole d'évaluation du site

L'évaluation est faite avec 25 questions posées à la PPIV après un moment non limité de lecture du site. Quatre personnes ayant différents degrés de perception visuelle ont participé à l'évaluation de notre site: une personne avec la vision tubulaire, une personne avec la vision périphérique, une personne aveugle de naissance, une personne déficiente visuelle tardive. Les participants au test pouvaient utiliser un lecteur d'écran avec ou sans transcription braille (Fig. 6) ; pour les malvoyants, le site pouvait être agrandi à 200% sur un grand écran. Les participants donnaient une note sur 10 des différents points évalués (10 étant la meilleure note).

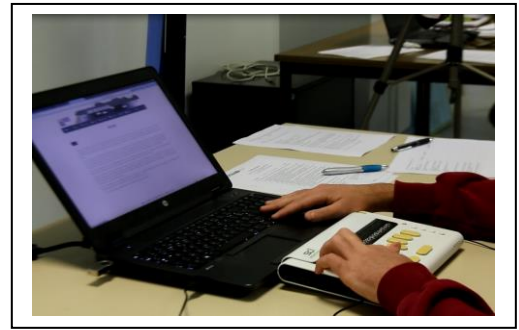

Fig. 6 Navigation sur le site Accesspace avec un lecteur d'écran et un transcripteur braille (plage tactile)

### E. Résultats d'évaluation préliminaire.

Le tableau 2 présente quelques critères d'accessibilité les plus importants parmi ceux considérés avec les PPIV, ainsi que les résultats collectés. L'évaluation a porté sur deux éléments : l'accessibilité à l'information (1) et la compréhension du contenu du site (2).

| Critère d'accessibilité | Accessible et indiqué avec le lecteur d'écran | Note /10 |
|---|---|---|
| 1. Liens | Indiqués | 10 |
| 2. Zones cliquables | Indiquer par un lien | 8,50 |
| 3. Images | Décrire mieux | 7,50 |
| 4. Vidéos | Pas d'audiodescription | 9,00 |
| 5. Changement de langue | L'anglais est lu comme le français | 8,25 |
| 6. Listes | Utiliser aussi pour la navigation | 6,67 |
| 7. Champs des formulaires | Indiquer le contenu exigé à l'intérieur du champ | 6,67 |
| 8. Abréviations | Indiquer la description | 5,00 |

Tableau 2. Evaluation de quelques critères d'accessibilité du site Accesspace par les participants à nos tests préliminaires.

1) *L'accessibilité à l'information* a été testée sur la navigation sur la page et dans le site ; et la recherche d'un onglet perdu. La possibilité de navigation sur le site (recherche des liens) avec le clavier ou la souris a été bien appréciée. La note moyenne est de 10/10, ce qui confirme la bonne place et l'annonce par le lecteur d'écran des balises associées aux liens.

Retrouver un onglet qui intéressait le participant (la recherche des « zones cliquables ») en cas de confusion sur le site, a été considéré comme facile et la note moyenne a été de 8,5/10. Cela confirme le bon emplacement de ces balises.

2) *Compréhension du contenu du site* : Le contenu des pages n'était pas toujours compréhensible avec le lecteur d'écran, surtout quand la configuration de l'outil n'était pas optimisée et personnalisée. En plus de l'information donnée par les images et les vidéos, leurs titres et l'audio (noté 7,5/10 pour les images et 9/10 pour les vidéos par rapport à leur description), les participants attendaient plus de description « visuelle » pour les images et l'audio description pour les vidéos. Les participants ont critiqué la synthèse vocale en anglais qui s'apparentait davantage à la lecture française d'un texte anglais. Les listes, les abréviations et les champs des formulaires ont été considérés insuffisamment décrits et lus trop rapidement ou de façon non intelligible (ex. abréviations).

Par ailleurs, la description de chaque page a été trouvée pertinente. Le bouton de retour à la page précédente n'était pas toujours repérable. Un texte trop long était difficile à survoler, les participants ont demandé de mettre des sections indiquant le contenu pour chaque paragraphe pour pouvoir lire plus vite.

Il est à remarquer que la spécificité de chaque participant a montré les besoins très différents des PPIV.

## VI. CONCLUSION ET PERSPECTIVES

Ce papier a proposé une analyse de l'accessibilité aux informations textuelles et visuelles sur internet et une méthode d'évaluation de l'accessibilité du site pour les PPIV. Il est nécessaire de proposer un standard de structuration logique d'un document lié à son contenu, guidé par une ontologie et par sa représentation interne arborescente, en tenant compte des mécanismes cognitifs de lecture et de la recherche de l'information.

Le développement des interfaces multimodales spécifiques pour chaque type de déficience et personnalisables semble être une priorité.

Pour la suite de cette étude préliminaire, il faudrait proposer un ensemble de programmes d'évaluation de sites (benchmark) qui teste différents paramètres comme la taille des caractères, les raccourcis, le codage de différentes informations, etc. Une enquête sur l'usage des lecteurs d'écran en France et en Francophonie [14] confirme la nécessité d'une normalisation et d'un programme d'évaluation de sites.